\documentclass[12pt]{article}
\usepackage{amssymb,amscd,amsmath,amsfonts}
\usepackage{amsthm,latexsym,feynmf,url}

\usepackage{hyperref}

\def\be{\begin{equation}}
\def\ee{\end{equation}}
\def\ba{\begin{array}}
\def\ea{\end{array}}

\newcommand{\D}{\mathcal{D}}
\newcommand{\Pd}[1]{\mathcal{P}^{(#1)}}
\newcommand{\N}{\mathbb{N}}

\newcommand{\V}{\mathcal{V}}
\newcommand{\scl}{\chi^{(d, 2)}_{[\frac{d}{2} - \ell, \mathbf{0}]}}
\newcommand{\spl}{\chi^{(d, 2)}_{[\frac{d+1}{2} - \ell, \mathbf{\frac{1}{2}}]}}

\newcommand{\chd}[1]{\chi^{(d)}_{(#1)}}

\newcommand{\z}{\mathcal{Z}}

\def\be{\begin{equation}}
\def\ee{\end{equation}}
\def\ba{\begin{array}}
\def\ea{\end{array}}

\def \v{{\vspace{0.75 cm}}}

\newlength{\blength}
\usepackage{color}
\definecolor{rougef}{rgb}{0.56,0,0}
\definecolor{vertf}{rgb}{0,0.5,0}
\definecolor{bleuf}{rgb}{0,0,0.8}

\begin{document}

\begin{center}
\textbf{{\LARGE Flato\,--\,Fronsdal theorem}} 

\vspace{.4cm}

\textbf{{\LARGE for higher-order singletons}}
\end{center}
\vspace*{.5cm}

\begin{center}
Thomas Basile$^{a,b,}$\footnote{E-mail address: \tt{thomas.basile@student.umons.ac.be}}, 
{Xavier Bekaert$^{a,}$\footnote{E-mail address: \tt{xavier.bekaert@lmpt.univ-tours.fr}}
and 
Nicolas Boulanger$^{b,}$\footnote{Research Associate of the Fund for 
Scientific Research-FNRS (Belgium); \tt{nicolas.boulanger@umons.ac.be}}
 }
\end{center}

\vspace*{.5cm}

\begin{footnotesize} 
\begin{center}
$^a$
Laboratoire de Math\'ematiques et Physique Th\'eorique\\
Unit\'e Mixte de Recherche $7350$ du CNRS\\
F\'ed\'eration de Recherche $2964$ Denis Poisson\\
Universit\'e Fran\c{c}ois Rabelais, Parc de Grandmont\\
37200 Tours, France \\
\vspace{2mm}{\tt \footnotesize }
$^{b}$
Service de M\'ecanique et Gravitation\\
Universit\'e de Mons -- UMONS\\
20 Place du Parc\\
7000 Mons, Belgique\\
\vspace*{.3cm}
\end{center}
\end{footnotesize}
\vspace*{.4cm}

\begin{abstract}

We prove a generalized Flato--Fronsdal theorem for higher-order, scalar and spinor, 
singletons. In the resulting infinite tower of bulk higher-spin fields, we point out the occurrence of 
partially-massless fields of all depths. 
This leads us to conjecture a holographic duality between a higher-spin gravity 
theory around $AdS_{d+1}$ with the aforementioned spectrum of  fields, and a free $CFT_d$ of 
fundamental  (higher-order) scalar and spinor singletons.
As a first check of this conjecture, we find that the total Casimir energy vanishes at one loop. 
\end{abstract}

\vspace*{.5cm}

\begin{center}
\textbf{\emph{\`A la m\'emoire de Francis A. Dolan}}
\end{center}
\newpage
\setcounter{footnote}{0}


\section{\large Introduction}

\v The celebrated Flato--Fronsdal theorem \cite{Flato:1978qz} is at the very core of the 
kinematics underlying Vasiliev's four-dimensional nonlinear field equations for higher-spin 
gravity \cite{Vasiliev:1988xc}.\footnote{For a non-technical review 
explaining the 
key mechanisms at work in Vasiliev's 4D equations, see \cite{Bekaert:2010hw}. For a 
recent technical review, see e.g. \cite{Didenko:2014dwa}.}
This theorem is also crucial for the higher-spin holographic conjectures \cite{Sezgin:2002rt,Klebanov:2002ja,Sezgin:2003pt} involving the free and critical 3-dimensional vector models of  
scalar and spinor fields  living at the conformal infinity of $AdS_{4}$.

\v From the purely group-theoretical point of view, 
the Flato--Fronsdal theorem can be stated as the decomposition of the tensor product of 
two Dirac singletons \cite{Dirac:1963ta} (either bosonic or fermionic) into an infinite direct sum of irreducible 
representations (irreps) of $\mathfrak{so}(2,3)\,$. 
The latter representations can either be viewed as on-shell Fronsdal fields 
propagating in the bulk of $AdS_{4}$ or as conserved currents of the $CFT_{3}\,$. 
Correspondingly, the Dirac singletons (scalar and spinor) can be viewed as $AdS_{4}$ 
fields whose physical degrees of freedom live  
at its conformal infinity or as free conformal fields  
propagating in 3-dimensional Minkowski space. 
The infinite sum of $\mathfrak{so}(2,3)\,$-irreps corresponds, from the bulk point of view, 
to an infinite tower of higher-spin gauge fields building up the spectrum of Vasiliev's 
theory. From the boundary point of view, the infinite sum corresponds to the single-trace 
sector of the corresponding vector models.

\v The Flato--Fronsdal theorem has been extended to higher-dimensional 
conformal algebras 
$\mathfrak{so}(2,d)\,$ in \cite{Heidenreich:1980xi,Vasiliev:2004cm,Dolan:2005wy}. 
In the work \cite{Vasiliev:2004cm}, 
the corresponding generalisations  suggest the existence of extended (supersymmetric) 
nonlinear higher-spin theories in $AdS_{d+1}$, where mixed-symmetry fields, 
on top of the totally-symmetric ones, are propagating, together 
with some massive $p$-forms. In the latter work \cite{Dolan:2005wy}, 
the whole analysis is based on characters  of $\mathfrak{so}(2,d)$ 
and incorporates singletons of spin $s\geqslant 1$ (for $d
$ even). 
Yet another class of generalisations features the so-called higher-order scalar 
singletons\footnote{Also called here multi-linetons in accordance with the terminology of 
\cite{Iazeolla:2008ix}  to which we refer for more details.}, 
that are non-unitary, ultra-short irreducible representations of the conformal algebra.
The spectrum of $AdS_{d+1}$ fields appearing upon 
tensoring out higher-order scalar singletons comprises partially massless gauge fields 
of odd depths \cite{Bekaert:2013zya}. Totally-symmetric partially-massless fields were introduced in 
\cite{Deser:1983tm,Deser:2001us} (the frame-like formulation 
was given in \cite{Skvortsov:2006at}) and their holographic duals are partially-conserved 
currents \cite{Dolan:2001ih}.
Gauge fields always turned out to be of crucial importance, thereby it can be suspected that
partially-massless field could be relevant in a unified framework for fundamental interactions, 
e.g. in a cosmological scenario \cite{Deser:2001xr} around $dS_{d+1}$  where 
they are unitary. 

\v In \cite{Bekaert:2013zya}, a version of higher-spin holography was 
proposed where a vector model at an isotropic Lifshitz fixed point is conjectured 
to be dual to a higher-spin gravity with spectrum given by a tower of massless 
and partially massless fields of unbounded integer spins -- see also \cite{Bekaert:2014} 
for a review of the conjecture. As we will explain below, we are lead to extend this conjecture to spinor CFT's. 

\v In the present note, we give a proof of the fusion rules presented in \cite{Bekaert:2013zya}
using character formulae and generalise them to include higher-order \emph{spinor} singletons. 
We actually compute the general case of the tensor product of two (scalar and spinor) singletons of 
different orders. A novelty
in the case of the tensor product of two higher-order \emph{spinor} singletons
is that the resulting spectra contain even-depth partially-massless fields. 
The spectra also contain mixed-symmetry fields that are massless, partially-massless
or massive.
Somewhat less expected, we find extra degeneracies in the spectrum 
of massless and partially-massless (both symmetric and mixed-symmetric) 
fields.\footnote{Mixed-symmetry, partially-massless fields seem to have been
first considered in \cite{Boulanger:2008up,Boulanger:2008kw}.}
Along the lines of the holographic conjectures proposed in \cite{Sezgin:2003pt} and 
\cite{Bekaert:2013zya}, 
we expect that the isotropic Liftshitz point of the Gross--Neveu model 
should be dual to a bulk theory containing the spectrum of fields found in the
present paper. 

\v If unbroken higher-spin gravity theories are indeed dual to free CFTs, they should therefore 
receive no quantum corrections. This is manifest in the non-standard off-shell 
proposal given in \cite{Boulanger:2011dd,Boulanger:2012bj}.
Several one-loop checks of this expectation were performed in 
\cite{Giombi:2013fka,Giombi:2014iua,Giombi:2014yra} 
where the total vacuum energy was shown to cancel.
As a first test of the validity of the holographic proposal we have made above with the 
isotropic Lifshitz fixed points, we find the cancellation of the total Casimir energy for all the spectra
considered in the present note.

\v The paper is organised as follows: 
After we explain, in Section~\ref{sec:Notation}, our methodology and notation, 
in Section~\ref{sec:characters} we compute the $\mathfrak{so}(2,d)$ characters of 
higher-order scalar and spinor singletons, 
as well as the characters of symmetric and mixed-symmetric fields, either (partially) massless or massive. 
The main tools employed in this section is a combination of the analyses presented 
in \cite{Dolan:2005wy} and \cite{Shaynkman:2004vu}.
In Section~\ref{sec:fusion}, we take the product of the characters of $\ell$th-order
scalar and spinor singletons found in the previous section and decompose the 
result into a sum of characters of indecomposable $\mathfrak{so}(2,d)$-representations. The 
central result of our paper is given in a theorem that generalises the Flato--Fronsdal 
theorem in the way explained above.
A corollary of our theorem is the computation, presented in Section~\ref{sec:Casimir}, 
of the total Casimir energy corresponding to the resulting spectra.
We conclude in Section~\ref{sec:conclusions} with some perspectives of future works. 
Finally, for the sake of completeness, in the Appendix we give a more general result 
compared to our theorem presented in 
Section~\ref{sec:fusion}, in the sense that it concerns products of singletons (scalar and 
spinor) of \emph{distinct} orders. The structure of the fusion rule is similar to the case 
presented in 
Section~\ref{sec:fusion}, but the depths of partially-massless fields run over 
different values and is more cumbersome to enunciate, which is the reason why we relegate 
that general result to the Appendix. This general case can fit with a Vasiliev-like higher-spin 
gravity construction by considering the tensor product of a \emph{direct sum} of distinct 
multi-linetons with itself.\footnote{Note that higher tensor products of 
singletons were considered in \cite{Engquist:2005yt,Engquist:2007pr} 
and in the Appendix E of \cite{Boulanger:2008kw}
where the non-compositeness of a generic Metsaev \cite{Metsaev:1995re} 
mixed-symmetry gauge field was shown.}

\section{\large  Notation and methodology}
\label{sec:Notation}

In this section, we spell out the approach that we followed in order to compute 
characters of $\mathfrak{so}(2,d)$ and their fusion rules. We also present our notation. 
The computation of the characters of the highest-weight representations we are 
interested in proceeds following a three-step procedure: 
\begin{itemize}
\item[(1)] We use the method explained in \cite{Dolan:2005wy} to compute the character 
of a Verma module. We then apply on it the $\mathfrak{so}(d)$ Weyl group symmetrizer 
in order to obtain the character of the corresponding \emph{generalised} Verma module;\footnote{For more 
details on the notions of Verma modules and generalised Verma modules relevant to our 
discussion, see e.g. \cite{Iazeolla:2008ix,Shaynkman:2004vu}.}
\item[(2)] The structure of possible (sub)singular modules of the resulting 
generalised Verma module is obtained 
upon inspection of the sequences given in the subsection 4.4. 
of  \cite{Shaynkman:2004vu}. 
It turns out that, for all the highest-weights we consider in detail, 
there is only one submodule which happens to be 
a generalised Verma module. The character of the latter is computed following the step 1;
\item[(3)] We finally subtract 
the characters obtained in the previous two steps in order to get the character of the corresponding irreducible module.  
\end{itemize}

\paragraph{Notation.}

The rank of the orthogonal subalgebra $\mathfrak{so}(d)\subset \mathfrak{so}(2,d)$ 
is denoted by $r\,$,  
\emph{i.e.} $D_r \cong \mathfrak{so}_\mathbb{C}(2r)$ and $B_r \cong \mathfrak{so}_\mathbb{C}(2r+1)$.  
In the present note, we will be only interested in $\mathfrak{so}(2,d)\,$ representations 
with highest-weight $\Lambda = (-\Delta , \vec{s}\,)\,$, where $\Delta$ is, in $AdS_{d+1}/CFT_d$ 
language, the scaling dimension of the corresponding primary operator or the minimal energy of the 
corresponding bulk field, and $\vec{s}\,=(s_1,\ldots,s_r)$ is a dominant-integral weight of 
$\mathfrak{so}(d)$ in the orthonormal basis.
More precisely, 
\begin{eqnarray}
s_1 \geqslant \ldots \geqslant s_r\geqslant 0 \quad \mbox{for}\quad d = 2r +1\;,
\nonumber \\
s_1 \geqslant \ldots \geqslant s_{r-1} \geqslant \vert s_r \vert \quad \mbox{for}\quad d = 2r\;,
\end{eqnarray}
where the $s_i$'s are either all integers or all half-integers. 
In the case where $d=2r\,$, the Dirac spinor can be decomposed into left and right irreducible Weyl 
spinors 
$(\tfrac{1}{2}\,, \ldots,\tfrac{1}{2}\,, \tfrac{1}{2}\,)$ and $(\tfrac{1}{2}\,, \ldots,\tfrac{1}{2}\,, - \tfrac{1}{2}\,)\,$,
and it will be denoted $\boldsymbol{\tfrac{1}{2}\,}$.

\v We denote by ${\cal V}(\Delta,\vec{s}\,)$ (resp. ${\cal D}(\Delta,\vec{s}\,)$) the generalised 
indecomposable (resp. irreducible) Verma module associated with the highest-weight  
$\Lambda = (-\Delta , \vec{s}\,)\,$. The character of ${\cal V}(\Delta,\vec{s}\,)$ is denoted 
by ${\cal A}^{(2,d)}_{[\Delta,\vec{s}\,]}(q,x_1,\ldots, x_r)$ while the character of 
${\cal D}(\Delta,\vec{s}\,)$
is denoted by ${\chi}^{(2,d)}_{[\Delta,\vec{s}\,]}(q,x_1,\ldots, x_r)\,$, where 
we used $x_{i} := e^{\mu_i}$ with a generic weight
$\vec{\mu} = \sum_{i=1}^r {\mu_i}\,\vec{\textsf{e}}_i \,$
in the orthonormal basis for $\mathfrak{so}(d)\,$. 
Similarly, $q := e^{-\textsf{e}_0}(\mu)=: e^{-\mu_0}$ where $\mu = \mu_0 \, \textsf{e}_0$ 
is a weight of $\mathfrak{so}(2)\subset\mathfrak{so}(2,d)\,$. The character of an irreducible representation of 
$\mathfrak{so}(d)\,$ with dominant-integral weight $\vec{s}\,$ is denoted by 
$\chi^{(d)}_{\vec{s}\,}(x_1,\ldots,x_r)\,$. 
We use the following shortening notation for the following specific cases:

\noindent $\bullet$ $(s_1,\ldots, s_j , \boldsymbol{s})$ is used to denote the following $\mathfrak{so}(d)\,$-weight 
$(s_1,\ldots, s_j , s,\ldots, s)\,$ for which 
\\
\noindent $s_{j+1}=s_{j+2}=\ldots = s_r=s\,$; 

\noindent $\bullet$ In the previous rule, when several last components of a weight are all null, 
we will sometimes omit the symbol $\boldsymbol{0}\,$. For example 
$(s_1,\ldots, s_j ,0,\ldots, 0)\,$ will sometimes be denoted $(s_1,\ldots, s_j )\,$ 
instead of $(s_1,\ldots, s_j ,\boldsymbol{0})\,$, for the sake of brevity;

\noindent 
$\bullet$ The weights $(s_1,\ldots, s_j , \underbrace{1,\ldots, 1}_{m} ,s_{j+m+1},\ldots, s_r)\,$ will be denoted 
$(s_1,\ldots, s_j , \boldsymbol{1}^m,s_{j+m+1},\ldots, s_r)\,$.

\section{\large Characters of partially-massless fields and higher-order singletons}
\label{sec:characters}

In this section we first review some results of  \cite{Dolan:2005wy,Shaynkman:2004vu} and 
use them to compute the characters of  partially-massless fields and higher-order singletons. 

\subsection{Review of general results}
\label{STV}

We recall the following two results spelled out in \cite{Shaynkman:2004vu}: 

\begin{itemize}
\item[(1)]
For $(\mu_1, \dots, \mu_r)$ 
  a $\mathfrak{so}(d)$ dominant-integral weight, and for
\begin{itemize}
\item {\underline{\it d = 2r +1:}}\\  
  If $\Delta$ is given by either $\Delta = r - \frac{1}{2} - n\,$ (for $n \in \N\,$) when 
  $\mu_1, \dots, \mu_r$ are all integers, or by  $\Delta = r - n\,$  (for $n \in \N\,$)
  when $\mu_1, \dots, \mu_r$ are all half-integers, then
  the irreducible module $\mathcal{D}(\Delta, \mu_1, \dots, \mu_r)$ is given by the following quotient:
  \begin{equation}
    \mathcal{D}(\Delta, \mu_1, \dots, \mu_r) \cong \frac{\V(\Delta, \mu_1,
      \dots, \mu_r)}{\D(d - \Delta, \mu_1, \dots, \mu_r)}\;,
    \label{quotient}
  \end{equation}
\item {\underline{\it d = 2r:}}\\  
If $\Delta$ is given by $\Delta = n - \mu_n$ for a given $\ n \in \{ 1,2,\ldots, r \}\,$, and obeys the 
conditions $\Delta \neq r$ and $-\Delta \pm \mu_r + r \in \N\,$, 
then the irreducible module $\mathcal{D}(\Delta, \mu_1, \dots,  \mu_r)$ 
is given by the same quotient as in~\eqref{quotient}.
\end{itemize}
In these two cases,  $\D(d - \Delta, \mu_1, \dots, \mu_r)\cong \V(d - \Delta, \mu_1, \dots, \mu_r)\,$, 
meaning that there is no singular sub-module in $\V(d - \Delta, \mu_1, \dots, \mu_r)\,$. 
\item[(2)] For $(\lambda_0, \lambda_1, \dots, \lambda_r)$ an $\mathfrak{so}(d + 2)$ 
dominant-integral weight, the irreducible module of highest-weight  $(-\lambda_1 - d + 1, \lambda_0 + 1, \lambda_2, \dots, \lambda_r)$
is given by:
  \begin{equation}
    \D(\lambda_1 + d - 1, \lambda_0 + 1, \lambda_2, \dots, \lambda_r)
    \cong \frac{\V(\lambda_1 + d - 1, \lambda_0 + 1, \lambda_2, \dots,
      \lambda_r)}{\D(\lambda_0 + d, \lambda_1, \lambda_2, \dots,
      \lambda_r)}
    \label{pmt}
  \end{equation}
with $\D(\lambda_0 + d, \lambda_1, \lambda_2, \dots, \lambda_r) \cong
  \V(\lambda_0 + d, \lambda_1, \lambda_2, \dots, \lambda_r)\,$. 
\end{itemize}

\subsection{Character of a generalised Verma module}

The character of an $\mathfrak{so}(2,d)$ Verma module is given by \cite{Dolan:2005wy}
\begin{equation}
  \mathcal{C}^{(d,2)}_{[\Delta,\vec{s}\,]} (q,x_1,\ldots,x_r) = q^{\Delta} \,\Pd d (q,x_1,\ldots,x_r) 
  \, \mathcal{C}^{(d)}_{\vec{s}\,} (x_1,\ldots,x_r)
  \label{char_mass}
\end{equation}
where 
\begin{equation}
\Pd d (q,x_1,\ldots,x_r) = 
\left\{ 
\begin{array}{c}
\frac{1}{(1-q)}\,  \prod\limits_{1\leqslant i \leqslant r} \frac{1}{(1- q x_i)(1 - q x^{-1}_i) } \quad \mbox{when}\quad d = 2r +1\;,
\\
\prod\limits_{1\leqslant i \leqslant r} \frac{1}{(1-q x_i)(1- q x^{-1}_i)} \quad\;\;\; \mbox{when}\quad d = 2r \;,
\end{array}\right.
\end{equation}
and 
\begin{equation}
\mathcal{C}^{(d)}_{\vec{s}\,} (x_1,\ldots,x_r) = 
\left\{ 
\begin{array}{c}
\prod\limits_{1\leqslant i \leqslant r}  x_i^{s_i+r+1/2 -i} (x^{{1}/{2}}_i - x^{-{1}/{2}}_i) 
\Delta^{-1}(x_1+x^{-1}_1,\ldots ,x_r+x^{-1}_r)  
~ \mbox{when}~ d = 2r +1\;,
\\
\prod\limits_{1\leqslant i \leqslant r}  x_i^{s_i+r-i} \Delta^{-1}(x_1+x^{-1}_1,\ldots ,x_r+x^{-1}_r)  
\quad\quad\quad\mbox{when}\quad d = 2r \;,
\end{array}\right.
\end{equation}
for $\Delta(x_1,\ldots,x_r) = \prod\limits_{1\leqslant i < j \leqslant r} (x_j - x_i)$ the Vandermonde determinant. 
It turns out that the rational function $\Pd d (q,x_1,\ldots,x_r)$ is invariant under the action of the Weyl group 
of $\mathfrak{so}(d)\,$.\footnote{The action of the Weyl group is reviewed e.g. in the Appendix B of \cite{Dolan:2005wy}. The Weyl group acts on the variables $x_i$ by permutations and changes of signs.}

\v In order to obtain the character of the corresponding \emph{generalised} Verma module, since the $\mathfrak{so}(d)$
part of the initial weight is always taken to be dominant integral, it suffices to apply the Weyl-group symmetrizer  
of $\mathfrak{so}(d)$ on 
$\mathcal{C}^{(d,2)}_{[\Delta,\vec{s}\,]} (q,x_1,\ldots,x_r)\,$, resulting in the following 
formula for the character of the corresponding generalised Verma module: 
\begin{equation}
\mathcal{A}^{(d,2)}_{[\Delta,\vec{s}\,]} (q,x_1,\ldots,x_r) = q^{\Delta} \Pd d (q,x_1,\ldots,x_r) 
  \chi^{(d)}_{\vec{s}\,} (x_1,\ldots,x_r)\;,
\end{equation}  
where, introducing 
\begin{equation}
k_i := s_i +r +\tfrac{1}{2}\,-i \;,
\end{equation}
the irreducible characters of $\mathfrak{so}(d)$ are  
\begin{equation}
\chi^{(d)}_{\vec{s}\,} (x_1,\ldots,x_r) = 
\left\{ 
\begin{array}{c}
\mbox{det}(x_i^{k_j} - x_i^{-k_j})\Delta^{-1}(x_1+x^{-1}_1,\ldots ,x_r+x^{-1}_r)
\prod\limits_{1\leqslant i \leqslant r} (x^{{1}/{2}}_i - x^{-{1}/{2}}_i) 
\\ 
~ \mbox{when}~ d = 2r +1\;,\quad \mbox{and}
\\
\\
\frac{1}{2}\,\left( \mbox{det}(x_i^{k_j} - x_i^{-k_j} ) + \mbox{det}(x_i^{k_j} + x_i^{-k_j} ) \right) 
\Delta^{-1}(x_1+x^{-1}_1,\ldots ,x_r+x^{-1}_r)
\\
\quad\quad\quad\mbox{when}\quad d = 2r \;.
\end{array}\right.
\end{equation}

\v

\subsection{Characters of various {specific} highest-weight modules}

\paragraph{Multi-linetons Di and Rac.}

Analogously to the terminology of Flato and Fronsdal, we call Di 
(resp. Rac) $\ell$-lineton the irreducible $\mathfrak{so}(2,d)\,$ 
module given by $\ell$ semi-infinite 
lines in compact weight space based on the lowest-energy 
states $\vert \tfrac{d+1}{2}-\ell , \boldsymbol{\tfrac{1}{2}}\rangle\,$
(resp. the state $\vert \tfrac{d}{2}-\ell , \boldsymbol{0}\rangle\,$).
Indeed, for $\ell = 1$ and $d=3$ they correspond to the Di (resp. Rac) singleton of 
Dirac \cite{Flato:1978qz}.
  
\v Using the results summarised in the subsection \ref{STV}, item (1), 
one finds that the irreducible modules corresponding to the Rac and Di $\ell$-linetons 
are given in terms of  the following quotients:
\begin{equation}
  {\cal{D}} \left(\frac{d}{2} - \ell, \mathbf{0} \right) \cong
  \frac{{\cal{V}}(\frac{d}{2} - \ell, \mathbf{0})}{{\cal{D}}(\frac{d}{2} + \ell,
    \mathbf{0})}
    \label{RAC}
\end{equation}
and
\begin{equation}
  {\cal{D}} \left(\frac{d+1}{2} - \ell, \mathbf{\frac{1}{2}}\right) \cong
  \frac{{\cal{V}}(\frac{d+1}{2} - \ell, \mathbf{\frac{1}{2}})}{{\cal{D}}(
    \frac{d-1}{2}+\ell, \boldsymbol{\tfrac{1}{2}})}
    \label{DI}
\end{equation}
with
 ${\cal{D}}(\ell + \frac{d}{2}, \mathbf{0}) \cong {\cal{V}}(\ell + \frac{d}{2},
\mathbf{0})$ and ${\cal{D}}(\ell + \frac{d-1}{2}, \mathbf{\frac{1}{2}}) \cong
{\cal{V}}(\ell + \frac{d-1}{2}, \mathbf{\frac{1}{2}})\,$. 
The characters of the corresponding modules are 
\begin{eqnarray}
  \scl(q,x_1,\ldots, x_r) &=& q^{\frac{d}{2}}(q^{-\ell} - q^{\ell}) \Pd d(q,x_1,\ldots, x_r)\;,
  \label{char_scl} \\
  & & \nonumber \\
  \spl(q,x_1,\ldots, x_r) &=& q^{\frac{d}{2}} (q^{- (\ell - \frac{1}{2})} - q^{\ell -
    \frac{1}{2}}) \chi^{(d)}_{\mathbf{\frac{1}{2}}}(x_1,\ldots, x_r) \Pd d (q,x_1,\ldots, x_r)\;.
  \label{char_spl}
\end{eqnarray}

\v In the case $d = 2r\,$, there is a restriction on $\ell$ for the modules to be irreducible.
It turns out that, for the scalar case, irreducibility implies $\ell<r\,$ and, for the spinor
case, $\ell\leqslant r\,$. This being said, when $\ell$ does \emph{not} 
satisfy these inequalities, the corresponding highest-weights are dominant integral. 
 In order to make the latter modules irreducible, one has to further quotient them and the 
resulting irreducible modules happen to be finite-dimensional. They correspond to 
$\mathfrak{so}(2,d)$ spherical harmonics, e.g. of spin (or degree) $\ell-r\geqslant 0$ for the scalar case
${\cal{D}} \left(r - \ell, \mathbf{0} \right)$. 
These irreducible subspaces of solutions of the higher-order D'Alembertian equations 
will not be of interest to us since they are finite-dimensional.
We will slightly abuse the notation and still use the symbols 
${\cal{D}} \left(\frac{d}{2} - \ell, \mathbf{0} \right)$ and 
${\cal{D}} \left(\frac{d+1}{2} - \ell, \mathbf{\frac{1}{2}}\right)$
for the corresponding \emph{reducible} quotient modules \eqref{RAC} and \eqref{DI}.

\v 
\paragraph{Bulk scalar fields.}

The structure of the 
module corresponding to a generic scalar field is
${\cal D}(\Delta, \mathbf{0})\cong {\cal V}(\Delta, \mathbf{0})\,$. 
The character for a scalar field propagating in the bulk of $AdS_{d+1}$ therefore is 
\begin{equation}
  \mathcal{A}^{(d,2)}_{[\Delta, \mathbf{0}]} (q,x_1,\ldots,x_r) = q^{\Delta} \Pd d (q,x_1,\ldots,x_r)\;.
  \label{char_mass_scalar}
\end{equation}
The unitarity condition is $\Delta \geqslant \frac{d-2}{2}\,$.  
The fields are reducible when $\Delta = \frac{d-2\ell}{2}\,$, for $\ell \in \mathbb{N}_0\,$, and 
for  these values of the conformal weight, one is back to the cases of the higher-order scalar singletons 
described above. 

\v With a slight abuse of language inherited from the philosophy in higher-spin gravity, 
the irreps ${\cal D}(d-1-t, \mathbf{0})\cong {\cal V}(d-1-t, \mathbf{0})$ will be called 
massless for $t=1$ and partially massless for $t>1\,$. 
This is justified by the fact that their scaling dimension 
$\Delta$ coincides with the scaling dimension of the corresponding spin-s irreps when 
$s=0$. However, as $\mathfrak{so}(2,d)$ irreducible modules, they have a different 
structure compared to the (partially) massless modules 
in the sense that they do not arise from a quotient of generalized Verma modules. 
Actually, this is not totally correct in even dimensions 
$d=2r\,$, where there exist values of $t$ for which ${\cal V}(d-1-t, \mathbf{0})$ admits singular 
sub-modules, so that the isomorphism ${\cal D}(d-1-t, \mathbf{0})\cong {\cal V}(d-1-t, \mathbf{0})$
is not true. 
This arises when $2r - 1 - t = r - \ell \Leftrightarrow t = r - 1 + \ell \,$ for $ \ell \geqslant 1\,$.

\v \paragraph{Totally-symmetric fields/operators.}
Using the results summarised in the subsection \ref{STV}, item (2), the structure of the 
modules corresponding to (partially-)massless, totally-symmetric fields/operators is 
\begin{equation}
  \D(s + d - t - 1, s) \cong \frac{\V(s + d - t - 1, s)}{\D(s + d - 1,
    s - t)}\;, \quad \mbox{for} \quad 1\leqslant t \leqslant s \;,  
\end{equation} 
where
$\D(s + d - 1, s - t) \cong \V(s + d - 1, s - t)$, 
whence the character:
\begin{align}
{{\cal{A}}^{(d, 2)}_{[s + d -t -1, s]}}(q,x_1,\ldots,x_r) & = q^{s + d - t - 1} \chi^{(d)}_{(s)}(x_1,\ldots,x_r)  
\Pd d(q,x_1,\ldots,x_r) & \text{when} \ t > s 
\\ {\chi^{(d, 2)}_{[s + d -t -1, s]}}(q,x_1,\ldots,x_r)  &
= q^{s + d - 1}\left( q^{-t} \chi^{(d)}_{(s)} - \chi^{(d)}_{(s - t)} \right) \Pd d(q,x_1,\ldots,x_r) 
  & \text{when} \ t \leqslant s\;.
  \label{char_pmt}
\end{align}
Totally-symmetric, spin-$s$, massless fields/conserved currents correspond to $t=1\,$. 
Totally-symmetric, depth-$t$, partially-massless fields/partially-conserved currents correspond to 
$2\leqslant t \leqslant s\,$. 
Totally-symmetric, massive fields/operators correspond to  $ t >s\,$. 

\v The same situation occurs for the modules 
$\V(s +d-t-1, s, \mathbf{\frac{1}{2}})$, $s$ half-integer, 
$\ t \in \N_0\,$,  
that possess one sub-module 
$\D(s + d -1, s - t, \mathbf{\frac{1}{2}}) \cong \V(s + d -1, s - t,\mathbf{\frac{1}{2}})$ for $s \geqslant t + 1/2\,$. 
We thus have:
\begin{equation}
  \D(s + d - t - 1, s, \boldsymbol{\tfrac{1}{2}}) \cong \frac{\V(s + d - t - 1,
    s, \mathbf{\frac{1}{2}})}{\D(s + d -1, s - t,
    \mathbf{\frac{1}{2}})}
\end{equation}
and the corresponding characters 
\begin{align}
  {{\cal A}^{(d, 2)}_{[s + d -t -1, s, \mathbf{\frac{1}{2}}]}}(q,x_1,\ldots,x_r) & 
  = q^{s + d - t - 1} \chd{s, \mathbf{\frac{1}{2}}}(x_1,\ldots,x_r) \Pd d(q,x_1,\ldots,x_r) &
  \text{when} \ t > s - 1/2 \\ 
   {\chi^{(d, 2)}_{[s + d -t -1, s, \mathbf{\frac{1}{2}}]}}(q,x_1,\ldots,x_r) & = q^{s + d - 1}\left( q^{-t} \chd{s,
    \mathbf{\frac{1}{2}}} - \chd{s - t, \mathbf{\frac{1}{2}}} \right)
  \Pd d(q,x_1,\ldots,x_r) & \text{when} \ t \leqslant s - 1/2\;.
  \label{char_pmtsp}
\end{align}

\paragraph{Mixed-symmetry fields/operators.}

Similarly, the modules with highest-weights 
$(t + 1 - s -d, s, \boldsymbol{1}^m)$ with $s, t \in \N_{0}$,  
possess one sub-module with highest-weight 
$(1 - s - d, s - t, \boldsymbol{1}^m)$ for  $s > t\,$. 
We thus have:
\begin{equation}
  \D(s + d - t - 1, s, \boldsymbol{1}^m) \cong 
  \frac{\V(s + d - t - 1, s, \boldsymbol{1}^m)}{\D(s + d - 1, s - t, \boldsymbol{1}^m)}
\end{equation}
with $\D(s + d - 1, s - t, \boldsymbol{1}^m) \cong \V(s + d - 1, s - t, \boldsymbol{1}^m)$. 
The characters are 
\begin{align}
{{\cal A}^{(d, 2)}_{[s + d -t, s + 1, \boldsymbol{1}^m]}}(q,x_1,\ldots,x_r)
   & = q^{s + d - t} \chd{s + 1, \boldsymbol{1}^m}(x_1,\ldots,x_r) \Pd d(q,x_1,\ldots,x_r) & \text{when} \ t
  \geqslant s \\ 
  {\chi^{(d, 2)}_{[s + d -t, s + 1, \boldsymbol{1}^m]}}(q,x_1,\ldots,x_r)
  & = q^{s + d}\left( q^{-t} \chd{s + 1, \boldsymbol{1}^m} - \chd{s
    - t + 1, \boldsymbol{1}^m} \right) \Pd d (q,x_1,\ldots,x_r)& \text{when} \ t < s \;.
  \label{char_pmtbm}
\end{align}
Hook-shaped, spin-$s$, massless fields/conserved currents correspond to $t=1\,$.\footnote{The mixed-symmetry gauge 
fields corresponding to $t=1$ were first discussed by Metsaev in \cite{Metsaev:1995re}.} 
Hook-shaped, depth-$t$, partially-massless fields or partially-conserved currents correspond to $2\leqslant t \leqslant s-1\,$. 
Hook-shaped, massive fields/operator correspond to  $ t \geqslant s\,$.

\v In the previous three cases, namely the totally-symmetric fields (bosonic and fermionic) 
and the mixed-symmetric ones, the massive modules can, in some sporadic cases, become 
reducible so that we abuse the notation when using the symbols $\cal A\,$ and $\cal D\,$.   
For example, in the totally-symmetric, massive spin-$s$ cases, when $ t >s\,$ and for $d=2r\,$, there are sporadic cases when the modules are reducible, in which 
cases we will abuse the notation when we will write $\D(s + d - 1, s - t)$ instead 
of $\V(s + d - 1, s - t)\,$.

\v \noindent
\paragraph{Unitarity.}
The modules 
$\D(s+d-t-1,s), \ \D(s+d-t-1, s, \mathbf{\frac{1}{2}})$ and
$\D(s+d-t, s, \boldsymbol{1}^m)$ are unitary only in the case 
$t = 1$, as are the scalar and spinor $\ell$-linetons $\D(\frac{d}{2}\,-\ell \,, \boldsymbol{0})$ and 
$\D(\frac{d+1}{2}\,-\ell\,, \boldsymbol{\tfrac{1}{2}})$ only in the case $\ell = 1\,$.

\v
\section{\large Fusion rules for higher-order singletons}\label{sec:fusion}

We now give the main result of the present note.

\v \textbf{Theorem.}
  \emph{The tensor product of two $\ell$-linetons, scalar and spinor, decomposes as:}
  \begin{equation}
    \D(\tfrac{d}{2} - \ell, \mathbf{0}) \otimes \D(\tfrac{d}{2} - \ell,
    \mathbf{0}) \cong \bigoplus_{k = 1}^{\ell} \bigoplus_{s = 0}^{\infty}
    \D(s + d - 2k, s,\boldsymbol{0})\;,
    \label{racrac}
  \end{equation}
  \begin{equation}
    \D(\tfrac{d+1}{2} - \ell, \mathbf{\tfrac{1}{2}}) \otimes
    \D(\tfrac{d}{2} - \ell, \mathbf{0}) \cong \bigoplus_{t = 1}^{2\ell - 1}
    \bigoplus_{s = \frac{1}{2}, \frac{3}{2}, \dots}^{\infty} \D(s + d
    - t - 1, s, \mathbf{\tfrac{1}{2}})\;.
    \label{dirac}
  \end{equation}
  
  \emph{In the cases  $d = 2r +1\,$, we have}
  \begin{align}
    \D(\tfrac{d+1}{2} - \ell, \mathbf{\tfrac{1}{2}}) & \otimes
    \D(\tfrac{d+1}{2} - \ell, \mathbf{\tfrac{1}{2}}) 
    \label{didi}
    \\ & \qquad\qquad  \cong
     \bigoplus_{t=2-2\ell}^{2\ell - 2}  \D \big(d-1-t, \mathbf{0}\big)   \oplus
    \bigoplus_{s=1}^{\infty} \bigoplus_{m=0}^{r-1}
    \D \left(s+d-2\ell, s, \mathbf{1}^m \right)\nonumber  
    \\  
    & \qquad\qquad\qquad\qquad \oplus 
    \bigoplus_{t=1}^{2\ell - 2} 
    \bigoplus_{s=1}^{\infty} \bigoplus_{m=0}^{r-1} 2 \D \left(
    s+d-t-1, s, \mathbf{1}^m \right) \;,
    \nonumber
    \end{align}
    
\emph{while in the cases $d=2r\,$,} 
\begin{align}
&    \D(\tfrac{d+1}{2} - \ell, \mathbf{\tfrac{1}{2}})  \otimes
    \D(\tfrac{d+1}{2} - \ell, \mathbf{\tfrac{1}{2}}) \cong 
 \bigoplus_{t=
 2-2\ell}^{2\ell-2 } 2\, \D \big(d-1-t, \mathbf{0}\big)   \oplus
   \nonumber \\ & \quad \oplus
    \bigoplus_{s=1}^{\infty} \bigoplus_{m=0}^{r-2}
   2\,  \D \left( s+d-2\ell, s, \mathbf{1}^m \right)  \oplus
   \bigoplus_{s=1}^{\infty}  
   \left[ \D(s+d-2\ell, s, \mathbf{1}^{r-2},1) \oplus \D(s+d-2\ell, s, \mathbf{1}^{r-2},-1) \right]    
   \nonumber \\
& \quad\oplus   \bigoplus_{t=1}^{2\ell - 2} 
    \bigoplus_{s=1}^{\infty} \bigoplus_{m=0}^{r-2} 4 \D \left( s+d-t-1, s, \mathbf{1}^m \right)  
 \nonumber \\
&   \quad\oplus \bigoplus_{t=1}^{2\ell - 2}\bigoplus_{s=1}^{\infty} \left[ 2 \D \left( s+d-t-1, s, \mathbf{1}^{r-2},1 \right) \oplus 
   2  \D \left( s+d-t-1, s, \mathbf{1}^{r-2},-1 \right) \right]\;.
   \label{DiDieven}
\end{align}
\\

\v\noindent We interpret these 3 decompositions as follows. 
\begin{itemize}
\item In the decomposition \eqref{racrac}, the tensor product of two Rac $\ell$-lineton 
gives \cite{Bekaert:2013zya} a sum over all integer-spin, partially massless, totally-symmetric fields 
of odd depths $t$ going from $t=1$ to $t=2\ell -1\,$, where strictly speaking the cases $t=1$ correspond 
to the \emph{massless} fields and the cases $t > s$ correspond to \emph{massive} fields;
\item In the decomposition \eqref{dirac}, the tensor product of a Di $\ell$-lineton with 
a Rac $\ell$-lineton gives a sum over all half-integer-spin, partially massless, 
totally-symmetric fields of all depths $t$ going from $t=1$ to $t=2\ell -1\,$;
\item In the decomposition \eqref{didi}, three types of fields appear:
\begin{itemize}
\item a finite multiplet of scalar fields, among which we find $2\ell-1$ massive fields
${\cal D}(d-1-t,\boldsymbol{0})$ for $2-2\ell\leqslant t \leqslant 0$ and $2\ell-2$ partially 
massless for $0<t\leqslant 2\ell-2$;
\item a finite collection $\{\D \left(s+d-2\ell, s, \mathbf{0} \right)\}_{s=1,2,\ldots,2\ell-2}$ of
massive spin-s fields;
\item an infinite collection $\{\D \left(s+d-2\ell, s, \mathbf{0} \right)\}_{s=2\ell-1,\ldots}$ of
partially-massless spin-s fields of depth $t=2\ell-1$;
\item a finite collection $\{\D \left(s+d-2\ell, s, \mathbf{1}^{m} \right)\}_{s=1,2,\ldots,2\ell-1; \ m = 1, \ldots, r-1}$ 
of mixed-symmetry (hook) massive fields;
\item an infinite collection $\{\D \left(s+d-2\ell, s, \mathbf{1}^{m} \right)\}_{s=2\ell,\ldots; \ m = 1, \ldots, r-1}$ of
partially-massless mixed-symmetry (hook) fields of depth $t=2\ell-1$; and 
\item a multiplicity-two tower of fields containing massive, partially-massless and massless
fields, either totally symmetric ($m=0$) or hook-shaped ($m>0$) of various depths below 
$2\ell-2\,$.   
\end{itemize}
\item In the even-dimensional case  the product of two Di $\ell$-linetons 
\eqref{DiDieven} differs from the odd-dimensional case \eqref{didi} by the overall multiplicity
of two, apart from the hook-like mixed-symmetry fields for which the first column has maximal height
$r\,$. We can track back the origin of this multiplicity by remembering that 
we actually consider the tensor product 
of two Dirac, not Weyl, $\ell$-linetons. The product of two left (resp. right) $\ell$-linetons will produce 
hook fields with a number of boxes in the first column with the parity of $r$, 
the tallest hook field having chirality left (resp. right).
The product of a left (resp. right)  Di $\ell$-lineton with a right (resp. left) Di $\ell$-lineton gives 
the hook fields with a number of boxes in the first column with the parity of $r+1\,$. 
\end{itemize}

\paragraph{Remark.} 
The proof of the above theorem amounts to decomposing a product of characters, 
namely rational functions of variables $q, x_1,\ldots, x_r\,$, into a sum of characters.  
The method of characters does not allow one to identify, 
in the aforementioned decomposition, 
whether the resulting characters are associated with 
decomposable or indecomposable modules. Since some of the representations 
considered here are non-unitary, they can be reducible but indecomposable, \emph{i.e.} 
not fully reducible.
Therefore, the symbols $\oplus$ appearing in the theorem should be understood in the weak sense, 
\emph{i.e.} either direct or semi-direct sums. Only an explicit realisation of the modules can allow the 
precise distinction, for example in a holographic context where the various modules are realised 
as fields with prescribed boundary behaviours.   

\v
\section{\large Casimir energy}\label{sec:Casimir}

Let us briefly recall the standard relations between characters of $\mathfrak{so}(2,d)$, 
canonical partition functions of $CFT_d$ and the Casimir energy on ${\mathbb R}\times S^{d-1}$ 
(see e.g. \cite{Giombi:2014yra,Gibbons:2006ij} for more explanations).

\v 
The canonical partition function ${\cal Z}(\beta)$ for a free boundary conformal field theory 
on ${\mathbb R}\times S^{d-1}$ (or for a free field in the anti-holographic bulk dual) 
in a thermal bath of temperature $T=1/\beta$ is equal to the corresponding 
$\mathfrak{so}(2,d)$-character evaluated at $q=\exp(-\beta)$ and $x_1=\ldots=x_r=1\,$.
The Hamiltonian zeta-function can be defined in terms of the Mellin transform of the canonical partition function
\begin{equation}
\zeta_E (z) :={\frac{1} {\Gamma(z)}\,} \int^\infty_0 d\beta\, \beta^{z-1} {\cal Z}(\beta)\,,
\label{zetaE}
\end{equation}
whose value at $z=-1$ is twice the Casimir energy $E_C= \frac12 \zeta_E(-1)$.

\v 
As pointed out in \cite{Giombi:2014yra}, the vanishing of the Casimir energy can sometimes be shown without 
computing explicitly the Hamitonian zeta-function due to the following property:
When the canonical partition function is an even function of the temperature, 
${\cal Z}(-\beta)={\cal Z}(\beta)\Leftrightarrow{\cal Z}(1/q)={\cal Z}(q)$, the Casimir energy automatically vanishes, 
$E_C=0\,$.
Let us shortly review the heuristic reasoning behind this fact.
As $\Gamma(z)$ has a pole at $z=-1\,$, the function $\frac{1}{\Gamma(z)}$ has a zero at $z = -1\,$, 
so for $\zeta_E(-1)$ to be non-zero, the integral has to have a pole at $z = -1\,$.
The expansion in Laurent series around $\beta=0$ of an even function ${\cal Z}(\beta)$ of course only contains even powers 
of $\beta\,$. Thus, the integral in \eqref{zetaE} can only have poles at even integer values of $z\,$. 
So the pole at $z=-1$ of the Gamma function at the denominator in \eqref{zetaE} cannot be compensated by a pole 
from the integral. Consequently, the Hamiltonian zeta-function vanishes at $z=-1$: 
$E_C=\frac{1}{2}\,\zeta_E(-1)=0\,$.

\v As we will show below, the canonical partition functions for any product of two (scalar or spinor) higher-order singletons 
happen to be even in $\beta\,$, thereby implying the vanishing of the total Casimir energy for the infinite 
tower of fields displayed in our theorem. 
More rigorously, we have checked this in two ways: On the one hand, we have computed 
the Hamitonian zeta function for the square of the $\ell$-lineton partition function; on the other hand,  
we computed the infinite sum of the individual Casimir energies. Both ways give zero.  

\v Using $\Pd d (q, 1 \dots, 1) = (1 - q)^{-d}\,$, we have the following expression 
for the partition function of a single Rac $\ell$-lineton:
\begin{equation}
  \z(q) = \scl(q, 1,\ldots, 1) = \frac{q^\frac{d}2(q^{-\ell} - q^\ell)}{(1 - q)^{d}}\;.
  \label{RaclPartition}
\end{equation}
This partition function $\z(\beta) $ is an even/odd function of $\beta$ when $d$ is odd/even:
\begin{equation}
  \z(q^{-1}) = \frac{q^{-\frac{d}2}}{(1 - q^{-1})^{d}} (q^{\ell} - q^{-\ell}) = (-1)^{d+1} \z(q)\,.
\end{equation}
The same parity  property holds for the partition function of a single Di $\ell$-lineton.
Therefore the partition functions for the tensor product of any two scalar or spinor higher-order 
singletons (possibly of distinct orders) are even functions of $\beta\,$, implying the vanishing 
of the corresponding Casimir energies.
One can check that this is indeed the case by a direct calculation. Here we show it for two 
Rac $\ell$-lineton, the calculation being similar for the other cases.

\v Using $\left( 1 - x \right)^{-\alpha} = \sum_{n = 0}^{+\infty} \binom{n + \alpha - 1}{\alpha - 1} x^n\,$, 
one finds that the Hamiltonian zeta-function of the square of $\z(\beta)$ given in \eqref{RaclPartition} is equal to
\begin{equation}
  \begin{aligned}
    \zeta_E(z) & = \frac{1}{\Gamma(z)}\int_0^{+\infty} d\beta \beta^{z-1} \frac{e^{-\beta d}(e^{2\ell \beta} + e^{-2\ell \beta} - 2)}{(1 - e^{-\beta})^{2d}} \\ & = \frac{1}{\Gamma(z)} \sum_{n = 0}^{+\infty} \binom{n + 2d - 1}{2d - 1} \int_0^{+\infty} d\beta \beta^{z-1} e^{-\beta (d + n)} 
    \left( e^{2\ell \beta} + e^{-2\ell \beta} - 2 \right) \;.
  \end{aligned}
\end{equation}
Now using $\int_0^{+\infty} d\beta \beta^{z-1} e^{-a \beta} = a^{-z} \Gamma(z)$, we have : 
\begin{equation}
  \zeta_E(z) = \sum_{n = 0}^{+\infty} \binom{n + 2d - 1}{2d - 1} \left[ (d + n - 2\ell)^{-z} + (d + n + 2\ell)^{-z} - 2(d + n)^{-z}\right]\;.
\end{equation}
For $z = -1\,$, the sum of the three terms in square brackets vanishes for any fixed value 
of $n\,$. As a result, $E_C = 0\,$.

\v Notice that, contrarily to the odd-dimensional case, the partition function of 
individual multi-linetons in even dimension $d$ are odd functions of $\beta$ 
and their Casimir energy are nonvanishing. Nevertheless, the Casimir energy vanishes for the tensor 
product of two of them. 

\v
%
\section{\large Conclusions and perspectives}\label{sec:conclusions}

In this note, we generalised the Flato--Fronsdal theorem to the cases of higher-order singletons, 
scalar and spinor, in arbitrary dimensions $d\geqslant 3\,$. 
We found novelties in the case of the tensor product of two spinor 
$\ell$-linetons, such as degeneracies and partially-massless mixed-symmetry 
fields of any depth. The tensor product of two scalar or spinor multi-linetons of any two different 
orders has been decomposed as well. This result is presented in the Appendix.

\v Our theorem in Section \ref{sec:fusion}
suggests the existence of a higher-spin, unbroken Vasiliev-like theory that should be 
the bulk dual of a Gross--Neveu model at an isotropic Lifshitz fixed point. 
The spectrum of this conjectured higher-spin gravity 
theory is encoded in the decomposition of two Di $\ell$-linetons given in our theorem.
A first check of this holographic conjecture and the one 
in \cite{Bekaert:2013zya} is the vanishing of the total Casimir energy at 1 loop for both spectra. 
These conjectures and their corresponding one-loop checks admit a straightforward generalisation 
for a direct sum of free scalar and spinor singletons of various orders.

\v An interesting generalisation of our results is to look at tensor products of higher-order spin-$s$ 
singletons for $s\geqslant 1\,$ (in even dimensions $d\,$).   
It would also be interesting to understand how the results obtained by the method of $\mathfrak{so}(2,d)\,$ characters 
can be ``continued" to the signature $\mathfrak{so}(1,d+1)\,$. This is motivated by the unitarity of partially-massless
fields in de Sitter geometries. For the same reason that the bosonic 
Vasiliev theory is unitary in de Sitter spacetime, 
the higher-spin gravity theories whose spectra we found in the present note should also be unitary in de Sitter background.    

 \v
\section*{\large Acknowledgments}

We are grateful to Ph.~Spindel for various discussions. The research of X.B. was supported by the Russian Science Foundation grant 14-42-00047 in 
association with Lebedev Physical Institute. X.B. wants to thank the 
F.R.S.-FNRS (Belgium) and the Service de M\'ecanique et Gravitation 
for financial support and hospitality during the completion of this work.
The work of N.B. was partially supported by a contract 
``Actions de Recherche concert\'ees -- Communaut\'e fran\c{c}aise de 
Belgique'' AUWB-2010-10/15-UMONS-1.

\begin{appendix}

\v
\section{General Theorem}

\v
  \emph{The tensor product of two higher-order singletons, scalar and spinor, decomposes as:}
  \begin{equation}
    \D(\tfrac{d}{2} - \ell, \mathbf{0}) \otimes \D(\tfrac{d}{2} - \ell',
    \mathbf{0}) \cong \bigoplus_{k = \tfrac{\vert \ell-\ell'\vert}2\,+1}^{\tfrac{\ell+\ell'}2}\bigoplus_{s = 0}^{\infty}
    \D(s + d - 2k, s,\boldsymbol{0})\;,
    \label{racracdiff}
  \end{equation}
  \begin{equation}
    \D(\tfrac{d+1}{2} - \ell, \mathbf{\tfrac{1}{2}}) \otimes
    \D(\tfrac{d}{2} - \ell', \mathbf{0}) \cong \bigoplus_{t = \big\vert \ell-\ell'-\tfrac12\big\vert+\tfrac12}^{\ell+\ell'- 1}
   \,\, \bigoplus_{s = \frac{1}{2}, \frac{3}{2}, \dots}^{\infty} \D(s + d
    - t - 1, s, \mathbf{\tfrac{1}{2}})\;.
    \label{diracdiff}
  \end{equation}
  
\noindent  \emph{In the cases  $d = 2r +1\,$, we have, for $\ell\neq \ell'\,$:}
  \begin{align}
   \D(\tfrac{d+1}{2} - \ell, \mathbf{\tfrac{1}{2}}) & \oplus 
   \D(\tfrac{d+1}{2} - \ell', \mathbf{\tfrac{1}{2}}) \cong 
   \nonumber \\ &
  \bigoplus_{t=\vert \ell - \ell'\vert }^{\ell + \ell'-2}  \left[ \D \big(d-1+t, \mathbf{0}
  \big) \oplus \D \big(d-1-t, \mathbf{0} \big) \right] \oplus
   \bigoplus_{s=1}^{\infty} \bigoplus_{m=0}^{r-1} \Big[ 
   \bigoplus_{t=\vert \ell - \ell'\vert +1 }^{\ell + \ell'-2} 2 \D \left(s+d-t-1, s, \mathbf{1}^m \right) 
   \nonumber \\ &
  \qquad \oplus   \D \left(s+d-\vert \ell - \ell'\vert -1, s, \mathbf{1}^m \right) 
   \oplus \D \left(s+d- \ell - \ell' , s, \mathbf{1}^m \right) \Big] \;,
    \label{didillprimeodd}
 \end{align}
    
\noindent \emph{while in the cases $d=2r\,$ and for $\ell\neq \ell'\,$:} 
\begin{align}
&    \D(\tfrac{d+1}{2} - \ell, \mathbf{\tfrac{1}{2}})  \otimes
    \D(\tfrac{d+1}{2} - \ell, \mathbf{\tfrac{1}{2}}) \cong
 \nonumber \\ &
  \bigoplus_{t=\vert \ell - \ell'\vert }^{\ell + \ell'-2}  \left[ 2 \D \big(d-1+t, \mathbf{0}
  \big) \oplus 2 \D \big(d-1-t, \mathbf{0} \big) \right] \oplus
   \bigoplus_{s=1}^{\infty} \bigoplus_{m=0}^{r-2} \Big[ 
   \bigoplus_{t=\vert \ell - \ell'\vert +1 }^{\ell + \ell'-2} 4 \D \left(s+d-t-1, s, \mathbf{1}^m \right)
 \nonumber  \\ & \qquad
   \oplus   2 \D \left(s+d-\vert \ell - \ell'\vert -1, s, \mathbf{1}^m \right) 
   \oplus 2 \D \left(s+d- \ell - \ell' , s, \mathbf{1}^m \right) \Big] 
  \nonumber \\ &  
   \oplus \bigoplus_{s=1}^{\infty} \Big[ 
   \bigoplus_{t=\vert \ell - \ell'\vert +1 }^{\ell + \ell'-2}\Big(  2 \D \left(s+d-t-1, s, \mathbf{1}^{r-2},1 \right)
   \oplus 2 \D \left(s+d-t-1, s, \mathbf{1}^{r-2},-1 \right)\Big)
   \nonumber \\ &
  \qquad \oplus    \D \left(s+d-\vert \ell - \ell'\vert -1, s, \mathbf{1}^{r-2},1 \right) 
   \oplus  \D \left(s+d- \ell - \ell' , s, \mathbf{1}^{r-2},-1 \right) \Big] \;.
    \label{didillprimeeven}    
\end{align}

\end{appendix} 

\providecommand{\href}[2]{#2}\begingroup\raggedright\endgroup


\end{document}